\documentstyle[12pt,aaspp4]{article}

\newcommand{\xte}{{\it RXTE}}
\newcommand{\psr}{1E~1048.1$-$5937}
\def\nudotdotdot{\ifmmode\stackrel{\bf \,...}{\textstyle \nu}\else$\stackrel{\,...}{\textstyle \nu}$\fi}
\def\dotdotdotnu{\ifmmode\stackrel{\bf \,...}{\textstyle \nu}\else$\stackrel{\,...}{\textstyle \nu}$\fi}
\def\lapp{\ifmmode\stackrel{<}{_{\sim}}\else$\stackrel{<}{_{\sim}}$\fi}
\def\gapp{\ifmmode\stackrel{>}{_{\sim}}\else$\stackrel{>}{_{\sim}}$\fi}

\begin{document}

\title{Long-Term \xte\ Monitoring of the Anomalous X-ray Pulsar \psr}

\author{
Victoria M. Kaspi,\altaffilmark{1,2,3}
Fotis P. Gavriil,\altaffilmark{1}
Deepto Chakrabarty,\altaffilmark{2}
Jessica R. Lackey,\altaffilmark{2} and
Michael P. Muno\altaffilmark{2}
}

\altaffiltext{1}{Department of Physics, Rutherford Physics Building,
McGill University, 3600 University Street, Montreal, Quebec,
H3A 2T8, Canada}

\altaffiltext{2}{Department of Physics and Center for Space Research,
Massachusetts Institute of Technology, Cambridge, MA 02139}

\altaffiltext{3}{Alfred P. Sloan Research Fellow}

\begin{abstract}

We report on long-term monitoring of the anomalous X-ray pulsar (AXP)
\psr\ using the {\it Rossi X-ray Timing Explorer}.  The
timing behavior of this pulsar is different from that of other AXPs
being monitored with \xte. In particular, we show that the pulsar
shows significant deviations from simple spin-down such that
phase-coherent timing has not been possible over time spans longer
than a few months.  We find that the deviations from simple spin down
are not consistent with single ``glitch'' type events, nor are they
consistent with radiative precession.  We show that in spite of the
rotational irregularities, the pulsar exhibits neither pulse profile
changes nor large pulsed flux variations.  We discuss the implications
of our results for AXP models. 
We suggest that
\psr\ may be a transition object between the soft gamma-ray repeater
and AXP populations, and the AXP most likely to one day undergo an
outburst.

\end{abstract}

\keywords{stars: neutron --- pulsars: general --- pulsars: individual (\psr) --- X-rays: stars}

\section{Introduction}

The nature of anomalous X-ray pulsars (AXPs) has been a mystery since
the discovery of the first example some 20 years ago. 
Although it is clear that they are young neutron stars, it is not 
clear why they are observable: their X-ray emission defies
conventional explanations that rely on either accretion or rotation
power as the radiation energy source.  Although only five AXPs are
known, their origin is likely to be of great importance to our
understanding of the fate of massive stars and the basic properties of
the young neutron star population.

AXP observational properties can be summarized as follows
(e.g. \cite{ms95}): (i) they have spin periods in the narrow range
$\sim$6--12~s; (ii) they exhibit regular spin-down on time scales of
$10^3$--$10^5$~yr; (iii) they show no evidence for Doppler shifts due
to binary motion; (iv) they lie in the Galactic plane, implying youth;
(v) two appear to be located at the geometric centers of supernova
remnants (SNRs), also indicating youth; (vi) all have much softer
X-ray spectra than accretion-powered pulsars; (vii) they have no known
counterparts at any other energy range, including radio or optical
(but see below); and (viii) their X-ray luminosities greatly exceed
their inferred spin-down powers, hence rotation power is implausible.
For an excellent recent review of these objects, see Mereghetti
(1999).

Suggested models for AXPs fall into two broad categories.
One model proposes that AXPs are young, isolated, highly
magnetized neutron stars or ``magnetars'' (\cite{dt92a}).  In this
case, AXP spin-down is due either to magnetic dipole radiation as in
radio pulsars, which implies large surface magnetic fields,
$10^{14}-10^{15}$~G, or it could be due to angular momentum loss from
Alfv\'en wave emission following crust cracking due to stresses caused
by an even larger magnetic field (\cite{tb98,tdw+00,dun00}).  Similar
fields are inferred independently in the soft gamma repeaters (SGRs)
which, aside from their soft gamma-ray outbursts, exhibit AXP-like
X-ray pulsations in quiescence (\cite{kf93,td95,kds+98}).  In this
model, the AXP X-ray luminosities could come from the decay of the
large magnetic field (\cite{td96a}) or from enhanced thermal emission
from the young cooling neutron star (\cite{hh97}).  The magnetar model
does not presently explain all AXP properties, however; for example it
is unclear why all the observed rotation rates should fall in so
narrow a range.

The second model of AXP emission is that they are powered by
accretion.  Conventionally, accreting material would come from a
companion (\cite{ms95}), however the absence of Doppler shifts allows
only very low mass and compact companions unless all known
systems are seen fortuitously face-on (\cite{mis98,wdf+99}).  The
companion may have been disrupted leaving behind a fossil accretion
disk (\cite{vtv95,cso+95}), although neither scenario is consistent
with the apparent youth of these objects.  An alternative is that AXPs
are neutron stars with accretion disks of material leftover from the
supernova explosion (\cite{chn00,alp99,mlrh01,ch00}). However, very constraining
optical observations of one AXP, 1E~2259+586, detect no evidence for
any companion, nor evidence for an accretion disk (\cite{hkvk00}).
A recent detection of a possible very faint optical
counterpart to the AXP 4U~0142+61 also renders the presence of an
accretion disk problematic (\cite{hvk00}).

One way to distinguish between these classes of models may be through
timing observations.  In the magnetar model for AXPs, relatively smooth
spin-down should be expected, punctuated by occasional abrupt spin-up
or spin-down events or ``glitches,'' as well as low-level,
long-time-scale deviations from simple spin-down, or ``timing noise.''
Both phenomena are well known among young radio pulsars (e.g.
\cite{lyn96}), although their physical origins in magnetars may be
considerably different given the much larger inferred magnetic field.
Indeed large amounts of timing noise have been seen in SGRs (e.g.
\cite{wkf+00}).  However no extended spin-up should be seen in
the magnetar model.  Further, Melatos (1999) \nocite{mel99} suggested
that deformation of the neutron star due to a large magnetic field
would result in radiative precession, which would manifest itself
observationally as a long-term periodicity in timing data.

On the other hand, accretion power is usually associated with much
noisier timing behavior, which can be correlated with spectral,
luminosity, and pulse morphology changes.  In addition, some accreting
binary systems undergo extended ($\sim$years) episodes of spin-up,
although these generally seem to alternate with long intervals of
spin-down as well (\cite{bcc+97}).

Until recently, monitoring of AXPs produced only measured frequencies
at individual epochs separated by many months or years.  This method
of timing is not sensitive to spin irregularities on a wide range of
interesting time scales.  Such ``incoherent'' timing can offer no
insight into the spin behavior of the source between observations, nor
to the abruptness of apparent changes in spin-down rate.  For example,
although 1E 2259$+$586 and \psr\ have the longest recorded timing histories of
all AXPs (e.g. \cite{cm97,bsss98,pkdn00}), the histories are poorly sampled and
observed apparent deviations from simple spin-down have been given
very different interpretations (e.g. \cite{mel99,hh99,bss+00}).

Kaspi, Chakrabarty \& Steinberger (1999) [hereafter KCS99],
\nocite{kcs99} making use of the scheduling flexibility of the {\it
Rossi X-ray Timing Explorer}, showed that ``phase coherent timing,''
that is, long-term monitoring that permits absolute pulse numbering
between observations (a technique that has been used for radio pulsars
for many years), can successfully be applied to at least two AXPs.
This implies that AXPs can be extremely stable rotators: KCS99 showed
that for AXPs 1E~2259+586 and RXS J170849.0$-$400910, deviations from
a simple linear spin-down model described the rotational behavior to
within a few percent of the pulse period over years.  Such stability,
together with phase-coherent timing, allowed Kaspi, Lackey, and
Chakrabarty (2000) \nocite{klc00} to detect a sudden spin-up event
having all the hallmarks of a classical radio pulsar glitch in
RXS~J170849.0$-$400910.

Whether such glitches are ubiquitous in AXPs, and how their amplitude
distributions and frequencies compare to those well studied in radio
pulsars, are unknown.  Similarly, whether great rotational stability
during glitch-free intervals is common to all AXPs is also unknown, as
is the origin of deviations from simple spin-down that have been seen
in the long-term timing histories of 1E~2259+586 and \psr.

\psr\ is a 6.4~s X-ray pulsar discovered in {\it Einstein} observations
of the Carina nebula (\cite{scs86}).  It has been shown to have a soft
X-ray spectrum (e.g. \cite{opmi98} and references therein), like other
AXPs.  It exhibits no evidence for any binary companion, as no Doppler
shifts of the pulse period are seen (\cite{mis98}), and no optical
counterpart to a limiting magnitude of $m_V \sim 20$ has been detected
(\cite{mcb92}).  Occasional monitoring observations over more than 20
years show that the pulsar is spinning down, though significant
deviations from a simple spin-down model have been noted
(\cite{mer95,cm97,opmi98,pkdn00,bss+00}). The paucity of data thus far
has made it impossible to unambiguously identify the origin of the
deviations.

Here we report on our monthly {\it Rossi X-Ray Timing Explorer} (\xte)
monitoring of \psr\ in which we have attempted phase-coherent timing
like that achieved for 1E~2259+586 and RXS J170849.0$-$400910.  We
show that the spin behavior of \psr\ is qualitatively and
quantitatively different from those of the other two AXPs.  In
particular, in our observations,
\psr\ has exhibited rotation that is much less stable than
those of the other AXPs timed as part of our \xte\ program.  We also
show that \psr\ occasionally suffers rotational anomalies on short
time scales that cannot be described by simple spin-down laws,
classical glitches, or radiative precession.  Further, we use the same
\xte\ observations to search for pulse profile changes and/or pulsed
flux variations, and to check for correlations with timing behavior.
We find no such correlations.  In \S\ref{sec:obs} we describe our
observations, in
\S\ref{sec:results} we describe our analysis procedures and results,
and we discuss the implications of our results in
\S\ref{sec:disc}.

\section{Observations}
\label{sec:obs}

The observations we report on were made with \xte's Proportional
Counter Array (PCA; \cite{jsg+96}).  The PCA consists of 5 collimated
xenon/methane multianode proportional counter units (PCUs), each
having a front propane anti-coincidence layer.  The instrument is
sensitive to photons in the energy range $2-60$~keV from a
$\sim1^{\circ}$ field of view (FWHM), with a total effective area of
approximately 6500~cm$^{2}$.  Observations of 3--6 ks in length of 1E
1048.1--5937 were made on a monthly basis during 1996 November
24--1997 December 13 and 1999 January 23--2000 August 11.  
%The latter
%were obtained as part of our dedicated Guest Observer program while
%the former are from the public archive.  
In addition, we used a handful of archival observations from 1995 and
1996; these generally had longer integration times than the other data
sets.  We used the GoodXenonwithPropane data mode, which records the
arrival time (1~$\mu$s resolution) and energy (256 channel resolution)
of all unrejected events.  Due to the soft spectrum of the object, we
only analyzed events from the top xenon layer of each PCU.  In the
timing analysis, we further restricted the data set by including only
those events that fall within an energy range of 2--5.5~keV,
as this maximizes the signal-to-noise ratio of the pulsar, given the
spectral properties of the source and background.  Photon arrival
times at each epoch were adjusted to the solar system barycenter
and binned at 62.5~ms intervals.

Our monitoring strategy for this source is identical to that for two
other AXPs as described in KCS99.  At first, each time
series was epoch-folded using the best estimate frequency determined
initially from either a periodogram or Fourier transform, though later
folding was done using an approximate timing ephemeris.
Resulting pulse profiles were cross-correlated in the Fourier domain with a high
signal-to-noise template created by adding phase-aligned profiles from
previous observations.  We implemented a Fourier-domain filter by
using only the first 6 harmonics in cross-correlation.  The
cross-correlation produces an average time-of-arrival (TOA) for each
observation.  The TOAs are then fit to a polynomial using the pulsar
timing software package {\tt TEMPO}.\footnote{\tt
http://pulsar.princeton.edu/tempo}  Pulse phase $\phi$ at any time $t$
can be expressed as a Taylor expansion
\begin{equation}
   \phi(t) = \phi(t_0)  + \nu(t-t_0)  + \frac{1}{2} \dot{\nu} (t-t_0)^2 + \frac{1}{6} \ddot{\nu} (t-t_0)^3 + \frac{1}{24} \nudotdotdot (t-t_0)^4 + \dots,
\label{eq:taylor}
\end{equation}
where $\nu\equiv 1/P$ is the pulse frequency, and $t_0$ is a reference epoch.
Unambiguous pulse numbering is made possible by obtaining monitoring
observations spaced so that the best-fit model parameters have a small enough
uncertainty to allow prediction of the phase of the next observation
to within $\sim 0.2$.  Typically this necessitates two closely spaced
observations (within a few hours of each other) followed by one spaced
a few days later, and regular monthly monitoring thereafter, as long
as phase coherence can be maintained.

To minimize use of telescope time, our monitoring data consisted of
brief (usually 3~ks) snapshots of the pulsar.  These snapshots suffice
to measure TOAs to good precision.  However, for any one epoch, the
period, as determined by Fourier Transform or epoch-folding, had
typical uncertainty of $\sim 3$~ms, which is quite large by normal timing
standards.  Thus, our snapshot method of measuring TOAs can determine
spin parameters with extremely high precision {\it only when phase
coherence can be maintained}.  However when this is not possible,
little information is available about the timing behavior of the
source.

\section{Results and Analysis}
\label{sec:results}

\subsection{Timing}
\label{sec:timing}

\begin{figure}
\plotone{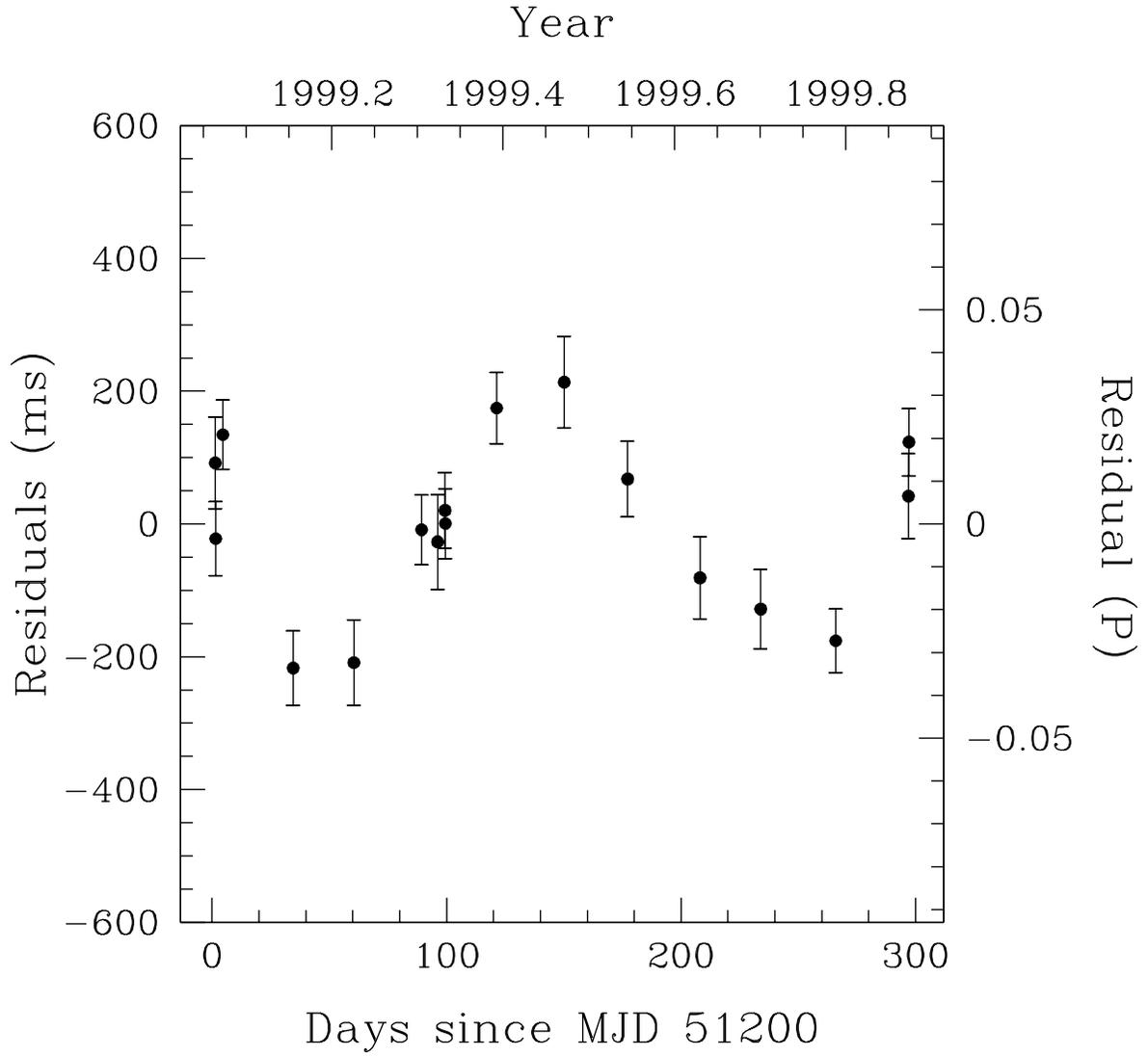}
\figcaption[f1.eps]{Arrival time residuals for \psr\ during 1999 with $\nu$, 
$\dot{\nu}$ and $\ddot{\nu}$ removed.  The remaining quartic trend is
clear; once it is fitted out by incorporating $\nudotdotdot$, the
residuals are trendless and consistent with zero within the
uncertainties.
\label{fig:res99}}
\end{figure}

We maintained unambiguous phase coherence for \psr\ in our monthly
observations from 1999 January 23 through 1999 November
15.  We required a fourth-order polynomial to characterize the 17
arrival times.  The best-fit $\nu$, $\dot{\nu}$, $\ddot{\nu}$ and
\nudotdotdot\ are listed in Table~\ref{ta:parms}; we call this fit
``Coherent 1999.''  The residuals (the differences between the
observed and model-predicted TOAs) are shown in Figure~\ref{fig:res99}
for the case when only $\nu$, $\dot{\nu}$ and $\ddot{\nu}$ are fitted;
the clear quartic trend shows that the $\nudotdotdot$ term is very
significant.  Following removal of the fourth-order polynomial, the
RMS residual is 49~ms, or 0.0075$P$, and the residuals are featureless.
%To demonstrate the importance of the terms in the fourth-order
%polynomial, consider their contributions to the phase via
%Equation~\ref{eq:taylor}.  The $\nu$ term contributes some 4 million
%phase turns, the $\dot{\nu}$ term 182, the $\ddot{\nu}$ term 17, and
%the \dotdotdotnu\ term 3.4.  Thus all are extremely significant (as is
%also verified by their fractional uncertainties -- see
%Table~\ref{ta:parms}).
These results alone clearly imply that the rotational behavior of
\psr\ is quite different from that of
1E~2259+586 and RXS~J170849.0$-$400910.  Those pulsars exhibit much
more stable rotation (apart from the one glitch in
RXS~J170849.0$-$400910) on comparable and even longer time scales,
that is, terms of higher order than $\dot{\nu}$ are very small or negligible for
those pulsars on time scales of over a year.  In fact, we show
next that \psr's rotation during the Coherent 1999 interval was
{\it more stable} than it was during other intervals spanned by our
observations.

After the observation of \psr\ on 1999 November 15, we were unable to
maintain phase coherence.  The next observation occured on 1999
December 12, 33 days later, and had phase residual 0.5 relative to our
Coherent 1999 fit (Table~\ref{ta:parms}).  So large a residual makes
absolute pulse numbering impossible.  Ascribing the entire anomaly to
a change in $\nu$, this implies a {\it minimum} change of $|\Delta \nu
/ \nu| \simeq 1 \times 10^{-6}$.  This is a lower limit since
additional full rotations could not have been detected.  The
subsequent three observations also could not be phase-connected so we
initiated a series of three closely spaced observations in
order to reachieve phase lock.  These took place starting
2000~April~11.  These unambiguously imply that the spin frequency,
averaged over the interval 2000~April~11--16 was
(0.15498892$\pm$0.00000013)~Hz, epoch 51650.0.  The 2000~April
observations and those obtained until the last epoch about which we
report (2000 August~11), have probably been properly phase connected;
the best-fit spin parameters are given in Table~\ref{ta:parms}.  We call
this fit ``Coherent 2000.'' It includes only $\nu$ and $\dot{\nu}$.
For the 8 TOAs included in the fit, the RMS residual was 198~ms or
0.08$P$, and no significant higher-order derivatives were necessary.
The high RMS in this interval suggests our phase-connection may not be
correct; if it is incorrect, then these 8 observations in all
probability cannot be connected at all, as the solution in
Table~\ref{ta:parms} is by far the best pulse numbering scheme that we
are able to find.

The frequency in the Coherent 2000 fit is {\it smaller} than that
predicted by the Coherent 1999 fit extrapolated to the Coherent 2000
epoch, by $\Delta\nu / \nu = (-1.25 \pm 0.08) \times 10^{-5}$
(assuming the frequency measured using the 2000 April
data alone).  This change is in the opposite direction to, and much larger
than that expected in a classical glitch, like those observed in
Vela-like radio pulsars as well as in RXS~J170849.0$-$400910,
although it is somewhat smaller than the possible ``anti-glitch''
observed in SGR 1900+14 (see
\S\ref{sec:disc}).  Furthermore, data from 1999 December through 2000
March (4 TOAs) do not fit with {\it either} the Coherent 1999 or
Coherent 2000 ephemerides.  Since those 4 TOAs are each separated by
one month, we cannot characterize the pulsar's behavior in that
interval.  However it is clear that the timing anomaly was not a
single classical glitch or even an ``anti-glitch,'' as the data cannot
be described by a single, sudden change in frequency.  If one wished
to characterize the pre-Coherent 2000 event as an anti-glitch, one
requires another event of comparable magnitude to have occured just
after the Coherent 1999 fit ended.

To investigate whether such behavior has occured in the past, we
analyzed archival \xte\ data obtained in 1995, 1996 and 1997.
The spacing of the available observations was
unfortunately non-optimal for a phase-coherent analysis.  However one
interval (1997 January 25 through 1997 May 18) had observation
spacings that we believe yield a reliable, phase-coherent
ephemeris in spite of potential phase counting errors.  This is
because, fortunately, {\it BeppoSax} observed the source on 1997 May
11, and we can use the observed frequency (\cite{opmi98}) to identify
the correct ephemeris by demanding consistency, as only one of the
possible solutions agrees.  We refer to this fit as ``Coherent 1997;'' it
is presented in Table~\ref{ta:parms}.

Although archival \xte\ observations at epochs within $\sim$1~month of
the start and end of the range spanned by Coherent 1997 exist, the fit
does not predict them well.  Thus it seems likely that, as in the
behavior we have observed in 1999--2000, the pulsar suffered some sort
of timing anomaly just prior to, and just after the Coherent 1997
span.  The alternative is that Coherent 1997 suffers from phase
counting errors and that the agreement with the {\it BeppoSax} data is
fortuitous, although we think that this is unlikely.

The spacing of the archival \xte\ data taken after the last epoch
spanned by the Coherent~1997 fit again has non-optimal spacing for a
phase-coherent analysis.  Furthermore, there are no
independent observations of \psr\ during this time that can help
discriminate among possible models.  However, our attempts at phase-coherent
timing show that only one model both characterizes the data well and
has $\dot{\nu} < 0$.  Although it is conceivable that during the short
span from 1997~July 17--December 13 the pulsar exhibited $\dot{\nu} >
0$, we assume that this is unlikely, given the strong evidence
against any spin-up in the 20~yr during which it has been observed.
This is admittedly dangerous as one goal of our observations is to
find epochs of spin-up; the assumption is therefore an important
caveat.  We refer to this fit as ``Coherent 1997a.''  The spin
parameters for this fit are given in Table~\ref{ta:parms}.

\begin{figure}
\plotone{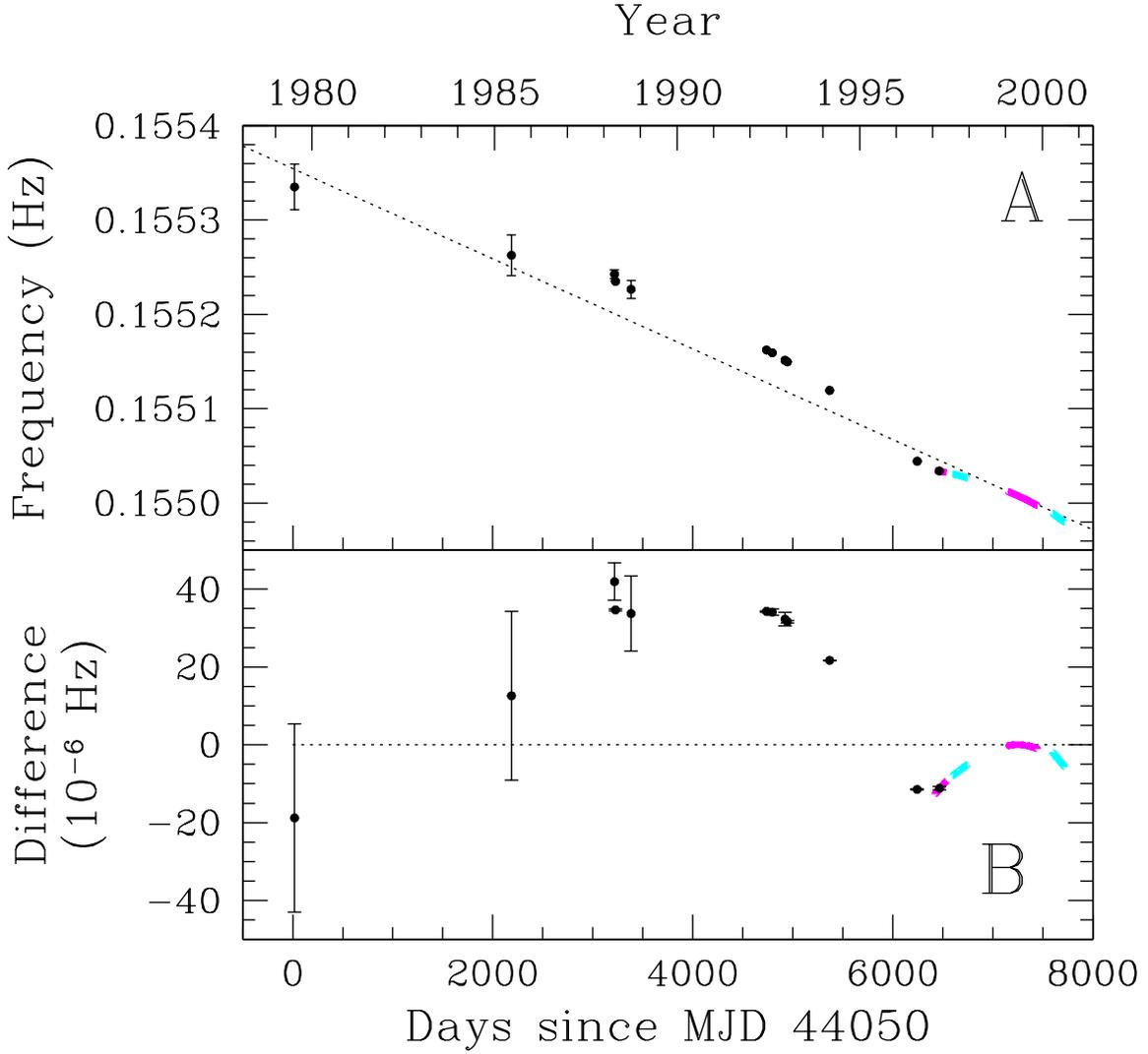}
\figcaption[f2.ps]{Spin history for \psr.  The points represent
past measurements of the frequency of the pulsar (see Oosterbroek et al. 1998 and 
references therein, Paul et al. 2000 and Baykal et al. 2000).  The solid lines
represent the phase-connected intervals as reported in this paper (see
Table~\protect\ref{ta:parms}).  
%The dashed line is a linear fit to the
%complete data set with a slope of
%-5.181911$\times$10$^{-8}\pm$9.00$\times$10$^{-14}$ and a frequency
%intercept of 0.155383$\pm$6.57$\times$10$^{-10}$ (CHECK).  
Panel A shows the observed frequencies over time.  The dotted
line is the extrapolation of the $\nu$ and $\dot{\nu}$ of the Coherent 1999 fit
(Table~\protect\ref{ta:parms}).  Panel B shows the
difference between the ephemeris indicated by the dotted line and the data points.
In both panels, the pink curves indicate the timing behavior during
intervals for which we have reliable phase-coherent ephemerides, while
the blue curve indicates a phase-coherent ephemeris that may have
pulse misnumberings (see \S\protect\ref{sec:timing} and
Table~\protect\ref{ta:parms}). \label{fig:history} }
\end{figure}

We can compare our pulse ephemerides with measurements of pulse
frequency made over the past 20~yr in order to look for long-term
trends.  Figure~\ref{fig:history} shows the resulting spin history of
\psr\ with previously measured spin frequencies plotted as points with
their corresponding 1$\sigma$ error bars.  Data were taken from the
compilation of Oosterbroek et al. (1998), \nocite{opmi98} with added
data from Paul et al. (2000) and Baykal et
al. (2000). \nocite{pkdn00,bss+00} Our timing ephemerides are plotted
as lines representing the four separate phase-connected segments (see
Table~\ref{ta:parms}).  Fits Coherent 1997 and 1999 are shown in
pink and are robust.  Fit Coherent 1997a and 2000 are shown in blue
because of the possibility that they may have pulse misnumberings (see
above).  In Figure~\ref{fig:history}, the dotted line represents an
extrapolation of the $\nu$ and $\dot{\nu}$ from the Coherent 1999 fit.
The lower plot shows the same data set with the linear term subtracted
off.  This magnifies deviations from the simple linear trend.
The long-term timing history clearly indicates that determining a
braking index will be difficult if not impossible as the long-term deviation
from linear spin down is not dominated by a simple $\ddot{\nu}$ term.

\subsection{Pulse Morphology}
\label{sec:morph}

Many accretion-powered pulsars are known to exhibit significant changes
in their average pulse profiles.  Such changes can be correlated
with the accretion state, and hence accretion torque and timing
behavior (\cite{bcc+97}).  
Furthermore, X-ray pulse profiles from the 
SGRs 1806$-$20 and 1900+14 have shown differences at different
epochs depending on time since outburst (\cite{kds+98,ksh+99}).
Such pulse profile changes, if they existed for \psr, would result 
in substantial timing anomalies, as our analysis assumes a fixed profile.  
To test this assumption, we have looked for evidence for pulse 
profile changes in our \xte\ monitoring data.

Pulse profiles were first phase aligned using the same
cross-correlation procedure used for timing.  Each profile was scaled,
a DC offset added, and the result subtracted from the average profile to
yield ``profile residuals'' for each observation.  To evaluate the
significance of the difference between the observed and average
profile, we calculated a $\chi^2$ statistic.  The resulting $\chi^2$
values were then compared with their expected distribution given the
number of degrees of freedom ($N-4$ where $N=64$ is the number of phase
bins) using a Kolmogorov-Smirnov (KS) test.  For \psr, the KS test
showed that the distribution of $\chi^2$ values is consistent with
having been drawn from a $\chi^2$ distribution having 60 degrees of
freedom.  Thus, \psr\ does not exhibit any significant pulse
profile variations, and our assumption of a fixed profile is
justified.

\subsection{Spectroscopy and Flux}
\label{sec:spectral}

In accreting systems in which the neutron star is undergoing spin-up,
changes in torque should be correlated with changes in X-ray flux,
since $\dot{\nu} \equiv N/2\pi I \propto \dot{M} \sqrt{r_m}$, where
$N$ is the torque, $I$ is the moment of inertia, $\dot{M}$ is the mass
accretion rate, and $r_m$ is the magnetospheric radius.  Accretion
torque theory predicts that $r_m \propto \dot{M}^{-2/7}$, and since
X-ray luminosity $L_x = GM\dot{M}/R$, where $M$ and $R$ are the
neutron star mass and radius, respectively, one expects $\dot{\nu}
\propto \dot{M}^{6/7}$ or $\dot{\nu} \propto L_x^{6/7}$
(\cite{pr72,lpp73}).  Though this naive prediction does not hold
precisely for accreting binaries, in general a strong correlation is
seen (\cite{bcc+97}).  For AXPs, because they are spinning down, the
above may not hold, however.  Chatterjee et al. (2000) suggest that
AXPs might be spinning down in the propeller regime due to accretion
from a fall-back disk.  In that case, $L_x \propto \dot{M}$ is still likely
to hold, along with $\dot{\nu} \propto \dot{M} r_m^2$ (see, e.g., Menou
et al. 1999).  Thus, if we take $r_m \propto \dot{M}^{-2/7}$ as above, 
then we expect $L_x \propto \dot{\nu}^{7/3}$, an even stronger correlation 
than in the simple spin-up case.  Since for \psr\ we find clear variations 
in $\dot{\nu}$, it is therefore interesting to ask whether there are 
correlated spectral and/or flux changes.

As pointed out by Baykal et al. (2000), given the large field-of-view
of the PCA and that the bright nearby but unrelated source $\eta$ Carinae
exhibited large flux changes over the course of our observations,
direct flux measurements of \psr\ could not be made with our \xte\
data.  Instead, we have determined the pulsed component of the flux,
by using off-pulse emission as a background estimator.  This renders
our analysis insensitive to changes in the fluxes of other sources in
the field-of-view.

Data from each observing epoch were folded at the expected pulse
period as for the timing analysis, except for this portion of the
analysis, we used only 8 phase bins.  For each phase bin, we
maintained resolution of 128 spectral bins over the PCA range.  We
used the off-pulse interval as our background measurement for this
analysis; given the broad morphology of the average pulse profile,
only one phase bin could be used.  The background phase bin was
determined by cross-correlating each (spectrally summed) profile with
the average pulse profile.  The pulse profiles were then phase
aligned, so that the same off-pulse bin was used for background in
every case.  The remaining 7 phase bins were summed, and their
spectral bins regrouped using the {\tt FTOOL} {\tt grppha}, such that
no bin had fewer than 20 counts after background subtraction.
Energies above 10~keV were ignored in our analysis.  The regrouped,
phase-summed data sets, along with the background measurement, were
used as input to the X-ray spectral fitting software package {\tt
XSPEC}\footnote{http://xspec.gsfc.nasa.gov} (\cite{arn96}).  Response
matrices were created using the {\tt FTOOL}s {\tt xtefilt} and {\tt
pcarsp}.

For the spectral analysis, although Oosterbroek et al. (1998) have
shown that the spectrum of \psr\ demands a two-component model, our
limited statistics and hence spectral resolution do not allow a
meaningful two-component fit.  Therefore we have fit our data with a
simple power law.  In all fits, $N_H$ was held fixed at $1.54 \times
10^{22}$~cm$^{-2}$, the value found by Oosterbroek et al. when they
used a simple power-law model.  Thus, in our fits, the power-law index
and normalization were free to vary.  Our fits are largely
insensitive to the choice of $N_H$ since we detect photons with
energies greater than 2~keV only.

%Figure~\ref{fig:spectral}A (top panel) shows our time series of
%measured photon indexes for \psr.  The ranges spanned by our coherent
%timing ephemerides, hence the epochs at which timing anomalies
%occured, are indicated.  
%We find that our measured photon indexes are stable to within our
%($\sim$25\%) uncertainties.  Thus, in spite of the substantial
%variations in $\dot{\nu}$ and the other spin parameters
%(Table~\ref{ta:parms}), we find no evidence for large spectral changes in
%the source.

\begin{figure}
\plotone{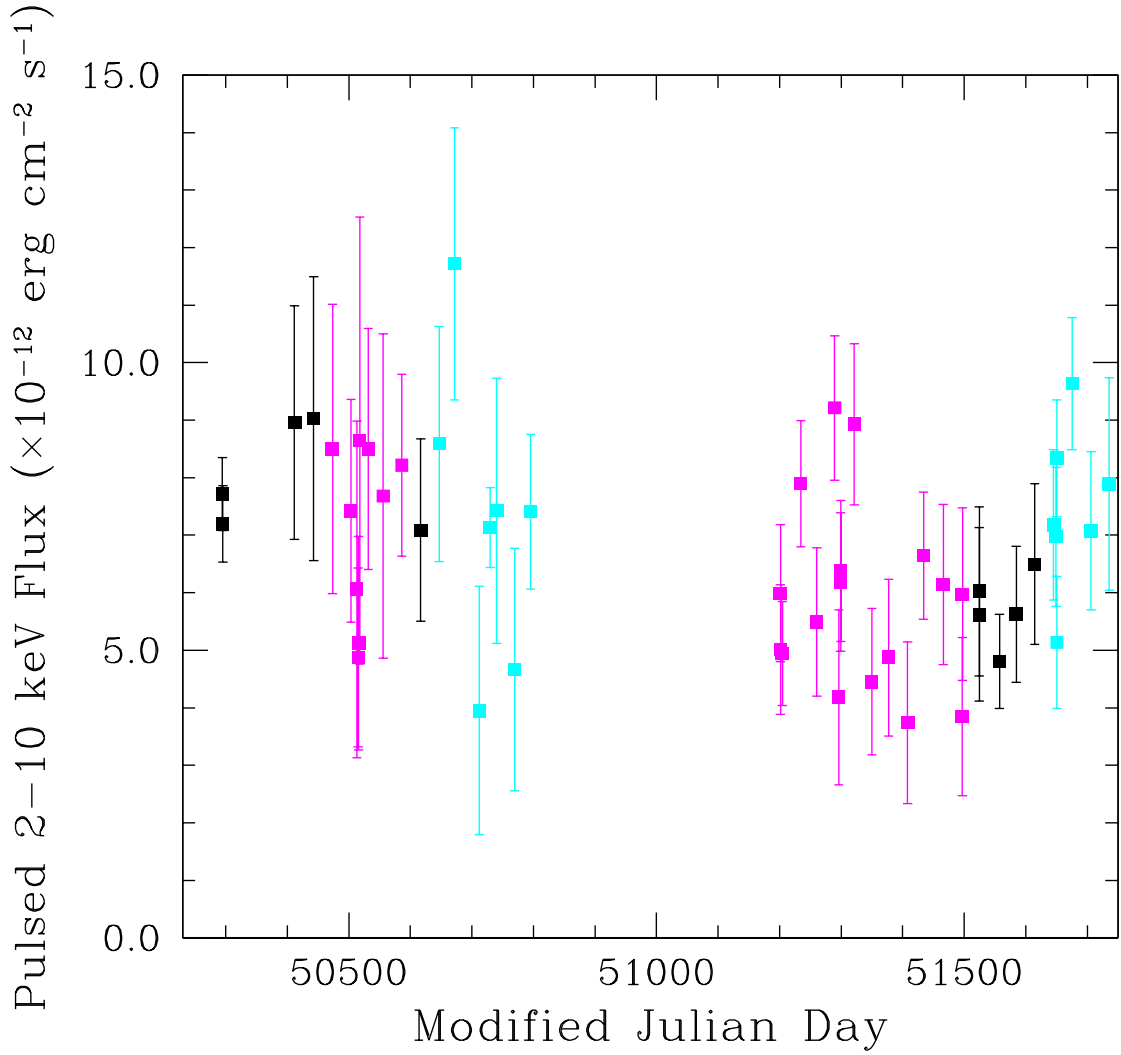}
\figcaption[f3.eps]{Pulsed flux time series in the 2--10~keV band
for \protect\xte\ observations
of \protect\psr.  The color coding is identical to that in
Figure~\protect\ref{fig:history}.  See \S\protect\ref{sec:spectral} for
details of the analysis procedure. \label{fig:flux}  }
\end{figure}

In order to extract a pulsed flux at each observing epoch, still using
{\tt XSPEC}, we reanalyzed each spectrum, this time holding the photon
index fixed at 3.9 (the mean photon index of all our observations --
we detected no significant variation in photon index within our
$\sim$25\% uncertainties in the parameter), as well as holding $N_H$
fixed at the same value as before.  This mean photon index in roughly
consistent, within uncertainties, with that found by Oosterbroek et
al. (1998) when they attempted a single-component model fit to their
{\it BeppoSAX} data.  At each epoch, only the normalization was fit,
and the uncertainty determined using the {\tt XSPEC} command {\tt
error}.  Corresponding 2--10~keV pulsed fluxes are plotted in
Figure~\ref{fig:flux}.  The color coding is identical to that given in
Figure~\ref{fig:history} and serves to indicate the epochs over which
we have phase-coherent timing solutions, and, more importantly, where
those epochs begin and end.  In particular, black points could not be
phase connected. It is clear that there are no significant changes in
pulsed flux associated with the epochs beginning and ending periods
over which we could phase connect.

In fact, we do not find evidence for any large variability in
the pulsed flux.  The mean 2--10~keV pulsed flux value is $(6.73 \pm
0.24) \times 10^{-12}$~erg~cm$^{-2}$~s$^{-1}$, the RMS flux value is
$2.1 \times 10^{-12}$~erg~cm$^{-2}$~s$^{-1}$ and the $\chi^2$ value
for deviations from the mean is 1.3 for 50 degrees of freedom.
This $\chi^2$ strictly speaking does suggest some low-level variability;
longer individual observations are clearly necessary to verify
this is the case.  However, as we discuss below
(\S\ref{sec:variability}), the pulsed flux is certainly much more
stable than previous analyses have suggested (Oosterbroek et al. 1998).

Note that our method results in fluxes that cannot be directly
compared with those made with instruments having higher spectral
resolution, as we have assumed an incorrect spectral model.  The same
is implied by our use of one phase bin for background, since imaging
instruments can use nearby blank fields for their background estimate,
and it is possible that some pulsar emission contaminates the
off-pulse bin.  Though our results may be systematically different
from those obtained with other instruments, the systematic shift
should be constant for all our \xte\ observations, implying that our
pulsed flux time series can be safely used to monitor the source's
pulsed flux variations.

\begin{figure}
\plotone{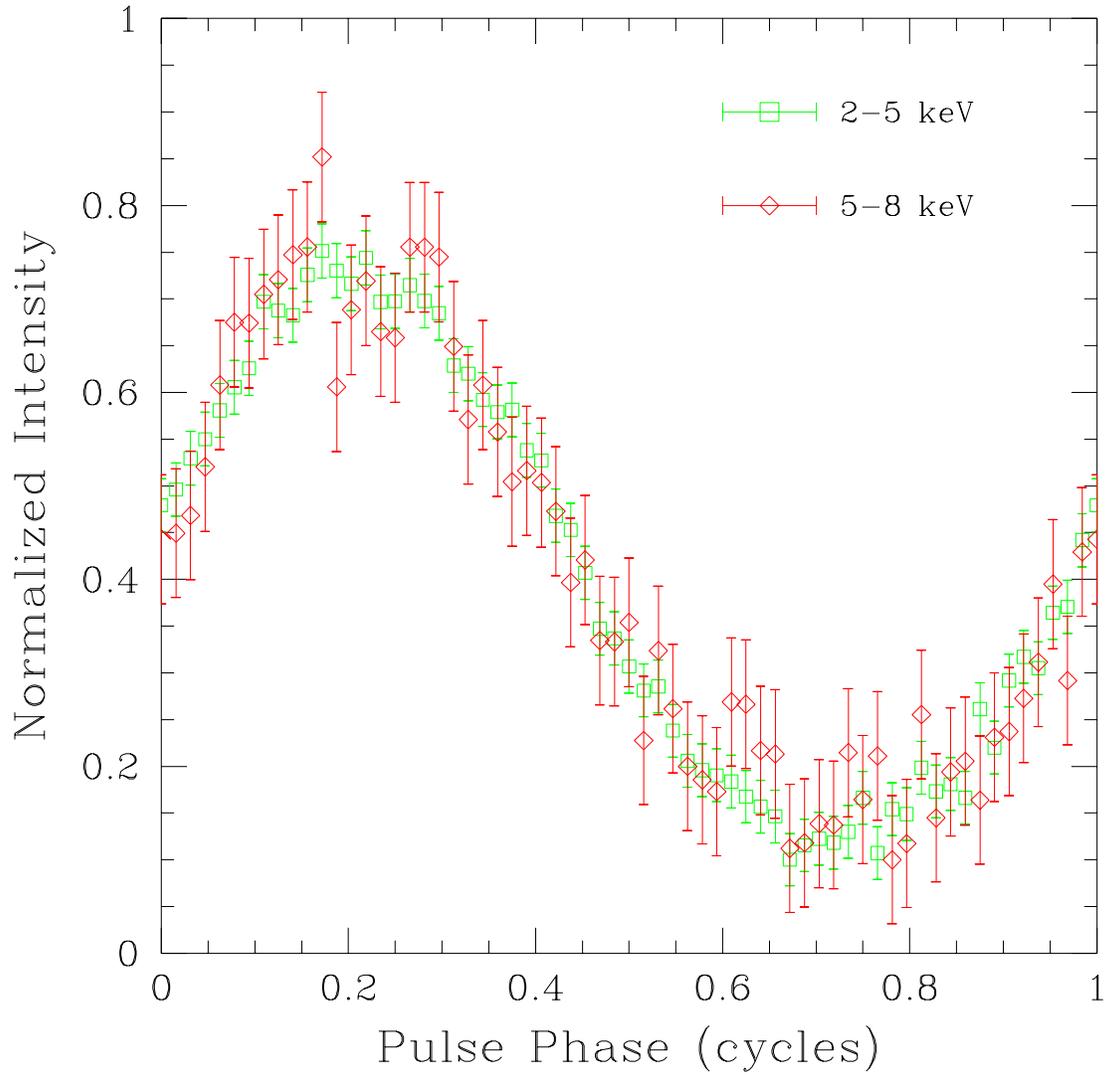}
\figcaption[f4.eps]{Average pulse profiles of \protect\psr\
in two energy bands as observed by \protect\xte.\label{fig:prof_energy} }
\end{figure}

Finally we note that the observed absence of spectral variations in
\psr\ is consistent with the observed absence of pulse morphology
variations (\S\ref{sec:morph}), because the pulse profile shows no
energy dependence.  This was verified by creating separate time series
in the energy ranges 2--5 and 5--8 keV, and folding each with the same
ephemeris, using the full 64 phase bins.  The resulting profiles were
scaled and had their DC level adjusted for comparison. The result is
shown in Figure~\ref{fig:prof_energy}, where it is clear that both
profiles have the same morphology to within the uncertainties.  This
is in contrast to what has been suggested for other AXPs
(\cite{ikh92,cso+95,wae+96,snt+97}), and to what we find from our
\xte\ observations of other AXPs (Gavriil \& Kaspi in preparation).
The absence of significant pulse profile changes with energy in \psr\
(see Fig.~\ref{fig:prof_energy}), even between energy ranges in
which thermal and non-thermal emission dominate, suggests that the two
mechanisms are related, or that the decomposition into two distinct
components is misleading.  This latter point is in contrast to what is
seen in rotation-powered pulsars, in which the thermal component
arises from cooling and results in a broad sinusoidal pulse, while the
non-thermal component is powered by the spin-down luminosity and
typically has a narrow pulse (e.g. \cite{shd99}).

\section{Discussion}
\label{sec:disc}

\subsection{Comparison of the Timing Properties of \psr\ with Those of Other Sources}

The timing behavior of \psr\ is different from that observed in the
two other AXPs for which phase-coherent timing results have been
reported, namely 1E~2259+586 and RXS~J170849.0$-$400910 (KCS99,
\cite{klc00}).  \psr\ has exhibited far noisier behavior on time
scales of weeks to months, to the point that phase-coherent timing
using monthly monitoring observations is in general not practical.
Denser sampling would certainly help.

Even at its most stable, \psr\ is much less stable a rotator than
other AXPs.  One method that has been used to characterize timing
noise in radio pulsars is the $\Delta_8$ statistic
(\cite{antt94}), where $\Delta_{8} \equiv
\log(|\ddot{\nu}|t^{3}/6\nu)$, for $t=10^{8}$~s.  Although not a
perfect characterization of stability, since it only accounts for the
phase deviation due to the second derivative as measured over a
single, specified time interval, Arzoumanian et al.  found that
$\Delta_8 = 6.6 + 0.6 \log \dot{P}$ provided a reasonable description
of their radio pulsar sample, with scatter about the relation of
approximately unity.  In that study,
$t=10^{8}$~s was chosen for consistency among sources, since
$\ddot{\nu}$ changes with time for random processes like the phase
and/or frequency wanderings seen in radio pulsars.  Measuring
$\ddot{\nu}$ for $t < 10^8$~s should, if anything, underestimate the
true value of $\Delta_8$.

For \psr, the longest phase-coherent
stretch we have is that described by the ``Coherent 1999'' fit
(see \S\ref{sec:timing}), which spans only $\sim$300 days, much
less than $10^8$~s.  Nevertheless, for this span $\Delta_8 = 3.8$,
while the Arzoumanian et al. relation predicts 0.2.  This clearly
implies much larger timing noise in \psr\ than is ever seen
in the radio pulsar population.  For comparison, the $\Delta_8$
values for the AXPs 1E~2259+586 and RXS~J170849.0$-$400910 are
consistent with those of the radio pulsar population (KCS99),
although the glitch in the latter must be accounted for.

As discussed in \S\ref{sec:timing}, single glitches in \psr\ cannot
explain our timing results because there exist data, for example, in the
interval between the Coherent 1999 and 2000 spans that do not fit
with either ephemeris.  This requires that two substantial glitches
have occured within three months.  Although such frequent glitching has
been seen in the radio pulsar population (e.g. in PSR B1737$-$30;
\cite{sl96}), such closely spaced glitches always have very different
amplitudes, with one typically an order of magnitude smaller than the
other.  This is presumably because it takes time to build up a
sufficient angular velocity differential between crust and superfluid
for there to be stresses on vortex lines that are large enough for a
major glitch to occur.  For \psr, both glitches would have had to have
been large, and, as noted in \S\ref{sec:timing}, must have been
``anti-glitches,'' ie sudden spin-down events.

Large deviations from simple spin down have been reported for SGRs
1806$-$20 and 1900+14.  SGR 1806$-$20 was timed phase-coherently over
a 178-day span, and showed considerable instability that was
characterized by a $\Delta_8$ value much greater than that of
radio pulsars, and comparable though somewhat larger than that
seen for \psr\ (\cite{wkf+00}).  As for SGR 1900+14, it exhibited
interesting timing behavior following a major outburst that occured on
1998 August 27 (\cite{mrl99,wkg+01}).  Because of sparse sampling
after the burst, it was not possible to distinguish between two
scenarios: (i) that the SGR suffered a major ($\Delta\nu / \nu \simeq
10^{-4}$) anti-glitch near or after the outburst, or (ii) that
following and not necessarily directly associated with the outburst,
it underwent a prolongued period of enhanced spin-down, with spin-down
rate a factor of $\sim$2.3 higher than the pre-burst rate.  The timing
behavior of
\psr\ is reminiscent of that of SGR 1900+14, as both sources have
shown substantial changes in spin-down rate, though the more frequent
monitoring of \psr\ has allowed us to rule out a single large
anti-glitch model.  However, the pulsed flux time series
(Fig.~\ref{fig:flux}) of \psr\ clearly shows that it exhibited no
outbursts or even enhanced pulsed flux at any time that could be
associated with its timing anomalies.  (We note that the pulsed
fraction of SGR~1900+14 did not change pre- and post- burst
(\cite{wkp+99}), therefore comparing its flux with the pulsed
component for \psr\ is reasonable.)  Furthermore, it has not shown any
pulse profile variations over the course of our observations, while a
substantial change in the X-ray light curve for SGR 1900+14 was
observed at the time of the outburst (\cite{ksh+99,wkg+01}).

\subsection{The Magnetar Model}

\subsubsection{Timing}

Mechanisms that have been invoked to explain the timing behavior of
SGR~1900+14 under the magnetar hypothesis, if they are to explain
\psr\ as well, must account for the differences in the behavior
of the two sources.  A large
``anti-glitch,'' a coupling between different components of the
differentially rotating magnetar via vortex line unpinning, has been
discussed to explain the event if it was sudden.  However, the glitch
trigger was most likely the large August 27 ouburst (\cite{tdw+00}).
Therefore this mechanism cannot explain the behavior of \psr.
Furthermore, as discussed above, no single glitch can describe the \psr\
data, and no SGR has been seen to undergo two major outbursts in so
short a time span.  Another model for the timing anomaly seen in SGR
1900+14 is that braking over a prolongued interval was driven by
particle outflow following a major, large-scale crust fracture
(\cite{tdw+00}).  However, this also demands a burst have occured, so
does not appear applicable to
\psr.  Thompson \& Blaes (1998) \nocite{tb98} have shown that
persistent seismic activity will accelerate the spin down of a neutron
star that is rotating slowly and has a large magnetic field.  Such
seismic activity, which generates Alfv\'en waves in the magnetosphere,
could be driven by small-scale crustal fractures, and variations in
the rate of these fractures could explain variations in spin-down
during outburst-free epochs (\cite{tdw+00}).  It remains to be seen
whether the variations in the spin-down rate of \psr, which are very
likely {\it larger} than those seen in the SGR, can be explained in
this model.  Duncan (2000) \nocite{dun00} has proposed that pure shear
oscillations, which manifest themselves as toroidal modes, could cause
angular momentum loss following crustal twisting fractures that need
not produce observable X-ray bursts if the bulk of the released energy
were in the deep crust.  He argues that the deviations from simple
spin-down seen in the long-term history of AXP 1E~2259+586 can be
accounted for in this scenario.  If so, then perhaps those of \psr,
though larger, can be as well.

Radiative precession, previously suggested to be the cause of
the long-term spin-down variations for
\psr\ (\cite{mel99}), appears to be ruled out by the now obvious
non-periodic nature of the frequency variations we have observed
(Fig.~\ref{fig:history}).  Indeed this model was already at odds with
the timing data for AXP 1E~2259+586 (KCS99), and continued
monitoring of the latter has only strengthened this conclusion (work
in preparation).  Radiative precession was predicted to occur
for neutron stars with magnetar-like magnetic fields, in which
rotation is like a rigid body.  However, such precession is likely
to be damped quickly because of the presence of superfluid and
vortex pinning in the neutron star interior (see \cite{klc00}
and references therein).
Thus, the absence of radiative precession is not inconsistent
with the magnetar model.

\subsubsection{Variability}
\label{sec:variability}

Our data (\S\ref{sec:spectral}, Fig.~\ref{fig:flux}) provide evidence, for the first time
using a single instrument and analysis procedure, that the pulsed flux
of \psr\ is relatively stable on time scales ranging from days to years,
although we did find marginal evidence for low-level ($< 25$\%)
variability.

Previously, Oosterbroek et al. (1998) compiled flux data from a
variety of different X-ray instruments that observed \psr.  That
compilation suggested that the pulsar shows variability by over a
factor of $\sim$5 on time scales of a few years, although the exact
factor is difficult to determine from their plot because of the
absence of uncertainty estimates.  The reality of those flux changes
is not supported by our results.  Given that
the various instruments have different spectral responses which could
result in inconsistent flux measurements if an erroneous spectral
model is used, it is perhaps not surprising that they find much
greater scatter than our data suggest exist.  One caveat is
that we measure pulsed flux, while they report flux, so the results
could be reconciled if the pulsed fraction is variable.  However the
pulsed fraction would have to be highly anti-correlated with the phase-averaged
flux in order to exactly cancel out variations in the pulsed flux.

%The time scale for variations that we observe, namely $\lapp$200~days,
%is not difficult to reconcile in the magnetar model.  The ultimate
%origin of the X-rays is thought to be the decay of the large stellar
%magnetic field and hence heat production deep in the neutron star
%interior.  The thermal conduction time from core to surface, estimated
%to be roughly a year for neutron stars having temperatures like that
%of \psr\ (\cite{rem91}) ought to smooth out variations on shorter time
%scales.  Thus, in the context of the magnetar model, our detection
%constrains the thermal conduction time.  On the other hand, that there
%exists a non-thermal component to the pulsar spectrum argues that some
%emission may come from the magnetosphere rather than the surface, in
%which case short term variations may be possible.  However, the
%absence of significant pulse profile changes with energy in this
%source (see Fig.~\ref{fig:prof_energy}), even between energy ranges in
%which thermal and non-thermal emission dominate, suggests that the two
%mechanisms are related.  This latter point is in contrast to what is
%seen in rotation-powered pulsars, in which the thermal component
%arises from cooling and results in a broad sinusoidal pulse, while the
%non-thermal component is powered by the spin-down luminosity and
%typically has a narrow pulse (e.g. \cite{shd99}).

Baykal \& Swank (1996) \nocite{bs96} found evidence for flux
variations by factor of up to $\sim$5 in 1E~2259+586 over the course
of 20 years, albeit also using data from many different instruments,
as in the Oosterbroek et al. compilation.
Baykal et al. (2000) have also reported a complete absence of
flux variations in this same source as measured with \xte\ over
$\sim$800~days.  They note that the steady flux coincides with epochs
over which the pulsar's timing has also been stable (KCS99) and
suggest that this is not a coincidence.  Our results for \psr\
demonstrate that the pulsed flux can be stable even in the presence
of significant torque variations, and also raise doubt about the
reliability of comparing fluxes made with different instruments.

%The absence of variations in flux for \psr\  support to the possibility that the
%AXP candidate AX~J1845.0$-$0300 is indeed an AXP.  Pulsations were
%detected only once for this source, following which its flux dropped
%by a factor of $\sim$10 (\cite{vgtg00}).  Such significant flux
%changes in AXPs would imply that there could be a large population of
%low-luminosity AXPs that could one day ``flare'' into observability.
%This would make it difficult to properly estimate the numbers and
%hence birth rates of such sources.

\subsection{Accretion Models}

%It is challenging to interpret the properties of \psr\ in terms of a
%binary accretion model.  
%As already discussed, \psr\ is quite
%different from the known accreting pulsars, as it has
%a very soft X-ray spectrum, lack of detectable Doppler shifts,
%absence of a detectable optical counterpart,
%lack of pulse shape variability, and steady pulsed flux.
Most of the known accreting pulsars are in high-mass X-ray binaries
with massive companions.  Given the absence of a luminous optical
counterpart in the AXPs including \psr, comparison to the low-mass
X-ray binary pulsars is more appropriate.  There are only five known
members of this class.  One of them, the 7.7~s accreting pulsar
4U~1626--67, shares many properties with the AXPs: its pulse period
falls in the same range, it shows no detectable Doppler shifts, it
underwent steady spin-down for years, and its X-ray luminosity is
steady for years at a time (\cite{cbg+97}).  However, 4U~1626--67 has
a known optical/UV counterpart (a 42-min binary period has been
inferred) and underwent an abrupt torque reversal accompanied by a
substantial change in X-ray spectrum (\cite{cbg+97}).  Neither before
nor after the torque reversal was the spectrum of 4U~1626--67 as soft
as AXP spectra (\cite{awn+95}).  
%Also, the hard X-ray spectrum
%of 4U~1626--67 contains a cyclotron absorption feature implying a
%$\sim 10^{12}$~G surface magnetic field (\cite{off+98}).

Nevertheless, it is of interest to compare the timing noise properties
of \psr\ with those of 4U~1626--67.  We computed a power spectrum of
the detrended (i.e. with $\nu, \dot{\nu}$, and $\ddot{\nu}$ removed)
phase residuals from the Coherent 1999 data using a Lomb-Scargle
periodogram (\cite{ptvf92}).  We quote torque noise strength in terms
of fluctuation power in pulse frequency derivative $P_{\dot\nu}$ (see,
e.g., \cite{boy81}).  Although our pulse phase measurements did not
span a sufficient time baseline to obtain a reliable measure of the
frequency dependence of the underlying noise process, it was possible
to constrain the noise strength over the observed time scales.  At
frequencies in the range $4\times 10^{-8} < f < 2\times 10^{-7}$ Hz
(i.e., time scales of 2--10 months), the power spectrum of
fluctuations in pulse frequency derivative $\dot\nu$ has a continuum
strength in the range (4--20)$\times 10^{-22}$
Hz$^2$~s$^{-2}$~Hz$^{-1}$.  This is much weaker than the level
observed in the persistent accreting pulsars (\cite{bcc+97}) with the
exception of 4U~1626--67, which has $P_{\dot\nu}\simeq 4 \times
10^{-22}$ Hz$^2$ s$^{-2}$ Hz$^{-1}$ \cite{cbg+97}).  We note that recent work
on the high-mass X-ray binary 4U~1907+09 shows that it too is
a comparably stable rotator (\cite{bia+01}).  Given that
in the Coherent 1999 interval \psr\ was at its {\it most} stable,
we conclude that the timing observations alone cannot rule out an
accreting binary scenario for this AXP.  Still, we regard
the case for \psr\ as an accreting binary, even with a very low-mass
companion, as weak, given the other evidence against this hypothesis,
namely, the very different spectrum, and the absence of pulsed flux or
pulse morphology changes correlated with the timing behavior, and the
steady spin-down over some 20~yr.

It is more difficult to dismiss the possibility that \psr\ is
accreting from a ``fallback'' disk comprising material left over from
the supernova explosion, since there is not yet a consensus on the
properties such a disk would have or on the expected timing and
variability properties of the pulsar.  However, one definite
expectation is that such a disk would be a significant emitter in the
optical and infrared, especially given the absence of a binary
companion to cause tidal truncation of the disk's outer edge.  The
present limits on optical/IR emission from \psr\ (\cite{mcb92})
already provide an interesting constraint.  Future optical/IR
observations following a more precise localization using the {\it
Chandra X-ray Observatory} should conclusively test the fallback disk
model.

\subsection{\psr: An AXP -- SGR transition object?}
\label{sec:transition}

The timing behavior of \psr\ appears to be different from that of
other AXPs on a variety of time scales.  While that of 1E~2259+586
also shows significant deviations from a simple spin-down model, this
pulsar has also shown tremendous stability on time scales of several
years, a characteristic which \psr\ has shown no evidence of
possessing.

Inspection of some of the overall properties of AXPs reveals that
\psr\ is unusual in other respects as well.  Table~\ref{ta:axps}
presents a compilation of spectral parameters and pulsed fractions for
the known AXP population (a similar compilation was presented by
Mereghetti 1999). \nocite{mer99} From the Table it is clear that \psr\
has the lowest photon index of any AXP (implying its non-thermal
component is the hardest) and the highest effective temperature
(implying its thermal component is the hottest).  Further, it shows
the highest ratio of blackbody to total
flux (once energy band is accounted for), and the largest pulsed fraction.

As discussed above, the timing instabilities of \psr\ are at least
qualitatively reminiscent of those seen for SGRs
(e.g. \cite{mrl99,wkf+00,wkg+01}).  We note that the harder spectrum
of \psr\ as characterized by its small photon index
(Table~\ref{ta:axps}) is in fact the closest of any AXP's to those
measured in the X-ray band for SGRs 1806$-$20 and 1900+14.  For those
sources, measured photon indexes are 2.2 in quiescence for SGR
1806$-$20 (\cite{smk+94}) and 1.1 pre-burst and 1.8 post-burst for SGR
1900+14 (\cite{wkg+01}), respectively.  Further, the thermal component
of \psr's spectrum has a high temperature (0.64~keV,
Table~\ref{ta:axps}) comparable to that seen for SGR 1900+14
post-burst, 0.62~keV (\cite{wkg+01}).  The pulsed fraction of
SGR~1900+14, however, only $\sim$11\% (\cite{wkp+99}), is considerably
lower than for \psr.

These similarities suggest that \psr\
may be a transition object between the AXP and SGR populations.  We
note that an AXP-like photon index recently measured for the quiescent
counterpart to SGR 0525$-$66 is steeper than those of the other SGRs,
and indeed is similar to those of the AXPs, providing
independent evidence for overlap of the two populations
(\cite{kkm+00}).  Thus, we suggest that \psr\ is the AXP that is the
most likely to one day exhibit SGR-like outbursts.  Continued \xte\
monitoring of \psr\ will easily detect such events; without such an
obvious signature, however, only more closely spaced monitoring
observations can determine the nature of the timing irregularities
exhibited by this ``anomalous'' anomalous X-ray pulsar.

\section{Conclusions}

Long-term \xte\ monitoring of the AXP \psr\ has shown it to be a
noisier rotator than the other AXPs monitored by \xte\ (KCS99).
Indeed, phase-coherent timing over spans longer than a few months has not been
possible for \psr.  Furthermore, we have observed deviations from
simple spin-down that are inconsistent with a single glitch event, and
inconsistent with being quasi-periodic as predicted for radiative
precession.  Only denser sampling will help determine the nature of
the spin-down irregularities in this source.  Our
\xte\ observations also show that the pulsar has exhibited no
pulse profile variations nor large pulsed flux variations.  The latter
is inconsistent with the simplest predictions of accretion theory,
although deep optical/IR upper limits are the best hope for
conclusively ruling out an accretion origin of the X-ray emission.
We note that the spin and spectral
properties of \psr\ are, among AXPs, the most reminiscent of the SGRs.
We therefore suggest that this pulsar may be a transition object
between the two populations, and the AXP most likely to one day
undergo an outburst.

\section*{Acknowledgements}

We are grateful to C. Thompson for useful discussions, and M. Roberts
and D. Psaltis for a critical reading of the manuscript.  This work
was supported in part by a NASA LTSA grant (NAG5-8063) and an NSERC
Research Grant (RGPIN228738-00) to VMK, with additional support from a
NASA ADP grant (NAG 5-9164).  This research has made use of data
obtained through the High Energy Astrophysics Science Archive Research
Center Online Service, provided by the NASA/Goddard Space Flight
Center. VMK thanks the Institute for Theoretical Physics in Santa
Barbara, CA for hospitality and acknowledges their partial support
through Grant No. PHY94-07194 from the National Science Foundation.

%\bibliographystyle{apj1c}
%\bibliography{journals1,modrefs,psrrefs,crossrefs}

\clearpage
\begin{deluxetable}{lcccc}
\tablecaption{Spin Parameters for \psr.\label{ta:parms} }
\footnotesize
\tablehead{
\colhead{} &  \colhead{Coherent 1997} &  \colhead{Coherent
1997a} & \colhead{Coherent 1999} & \colhead{Coherent 2000} }
\startdata
Dates & 1997 Jan -- May & 1997 Jul -- Dec & 1999 Jan -- Nov & 2000
Apr -- Aug \\
First observing epoch (MJD) & 50473 & 50646 & 51201 & 51645 \\
Last observing epoch (MJD) & 50586 & 50795 & 51497 & 51767 \\
Total number of observations & 11 & 10 & 17 & 8 \\
$\nu$ (Hz) & 0.155033822(3) & 0.155028540(3) & 0.1550076627(13) & 0.154988861(13) \\
$\dot{\nu}$ ( $10^{-13}$~Hz~s$^{-1})$ & $-$2.054(12) & $-$3.328(12) &
$-$5.5404(18) & $-$9.15(3)\\
$\ddot{\nu}$ ( $10^{-21}$~Hz~s$^{-2})$ & - & - & $-$6.18(10) & - \\
\nudotdotdot\ ( $10^{-28}$~Hz~s$^{-3})$ & - & - & $-$1.9(2) & - \\
Epoch of $\nu$ (MJD) & 50515.0 & 50730.0 & 51299.4 & 51650.0 \\
RMS residual (ms) & 93 & 156 & 49 & 198 \\
\enddata
\end{deluxetable}

\clearpage
\begin{deluxetable}{lccccc}
\tablecaption{Properties of Known Anomalous X-ray
Pulsars\tablenotemark{a}. \label{ta:axps}}
\footnotesize
\tablehead{
\colhead{NAME } & \colhead{$kT$ (keV)\tablenotemark{b}} & \colhead{$\Gamma$\tablenotemark{c}} &
\colhead{$L_{BB}/L_{tot}$\tablenotemark{d}} & \colhead{Pulsed Fraction \tablenotemark{e}} &
\colhead{Refs.} }
\startdata
4U 0142+61 & 0.34--0.44 & 3.4--4.2 & 0.2--0.38  (0.5--10 keV) &
13\% (4--10 keV) & 1, 2 \\
{\bf \psr\ } & $0.64 \pm 0.01$ & $2.52 \pm 0.20$ & 0.55 (2--10 keV) & 70\% (0.5--10
keV) & 3 \\
1E 1841$-$045 & \nodata & $3.4 \pm 0.3$ & \nodata & $\sim$15\% (1--10 keV) & 4, 5\\
RXS~J170849.0$-$400910 & $0.41 \pm 0.03$ & $2.9 \pm 0.3$ & 0.17 (0.8--10 keV) & $\sim$50\% (4--10 keV) & 6  \\
1E 2259+586 & $0.44 \pm 0.01$ & $3.93 \pm 0.09$ & $\sim$0.58 (1--10 keV)
& 35\% (1--10 keV) & 7, 8, 9 
\enddata

\tablenotetext{a}{The AXP candidate AX J1845.0$-$0300
(\cite{tkk+98,gv98}) is omitted pending confirmation.}
\tablenotetext{b}{Equivalent black-body temperature.}
\tablenotetext{c}{Power-law photon index.}
\tablenotetext{d}{Fraction of total luminosity in the black-body component.}
\tablenotetext{e}{The given pulsed fractions have been estimated using
a consistent definition so may differ from the
originally reported value -- see Pivovaroff, Kaspi \& Camilo
(2000).\nocite{pkc00}}

\tablerefs{
(1) \cite{wae+96}; 
(2) \cite{ioa+99};
(3) \cite{opmi98}; 
(4) \cite{gv97}; 
(5) \cite{vg97}; 
(6) \cite{snt+97};
(7) \cite{cso+95}; 
(8) \cite{rp97}; 
(9) \cite{pof+98} }

\tablecomments{All spectral parameters quoted are from two-component 
fits as reported in the given references, unless only a single
component fit was done.  Uncertainties are as quoted in the
references.  Where different observations conflicted (e.g. for
4U~0142+61), a range is given.}
\end{deluxetable}

\normalsize
\clearpage

\clearpage

\clearpage

\clearpage

\end{document}